\documentstyle[12pt,epsfig]{article}
\begin{document}

\def\lta{\;\raisebox{-.5ex}{\rlap{$\sim$}} \raisebox{.5ex}{$<$}\;}
\def\gta{\;\raisebox{-.5ex}{\rlap{$\sim$}} \raisebox{.5ex}{$>$}\;}
\def\grle{\;\raisebox{-.5ex}{\rlap{$<$}}    \raisebox{.5ex}{$>$}\;}
\def\legr{\;\raisebox{-.5ex}{\rlap{$>$}}    \raisebox{.5ex}{$<$}\;}

\newcommand{\ra}{\rightarrow}
\newcommand{\permille}{$^0 \!\!\!\: / \! _{00}$}
\newcommand{\dd}{{\rm d}}
\newcommand{\oal}{{\cal O}(\alpha)}%
\newcommand{\su}{$ SU(2) \times U(1)\,$}
 
\newcommand{\eps}{\epsilon}
\newcommand{\mw}{M_{W}}
\newcommand{\mww}{M_{W}^{2}}
\newcommand{\mbb}{m_{b \bar b}}
\newcommand{\mcc}{m_{c \bar c}}
\newcommand{\mbc}{m_{b\bar b(c \bar c)}}
\newcommand{\mh}{m_{H}}
\newcommand{\mhh}{m_{H}^2}
\newcommand{\mz}{M_{Z}}
\newcommand{\mzz}{M_{Z}^{2}}

\newcommand{\lra}{\leftrightarrow}
\newcommand{\tr}{{\rm Tr}}
 
\newcommand{\ie}{{\em i.e.}}
\newcommand{\cm}{{{\cal M}}}
\newcommand{\cl}{{{\cal L}}}
\def\Ww{{\mbox{\boldmath $W$}}}  
\def\B{{\mbox{\boldmath $B$}}}         
\def\nn{\noindent}

\newcommand{\sinsq}{\sin^2\theta}
\newcommand{\cossq}{\cos^2\theta}

\newcommand{\epem}{$e^{+} e^{-}\;$}
\newcommand{\epemt}{e^{+} e^{-}\;}
\newcommand{\eeah}{$e^{+} e^{-} \ra H \gamma \;$}
\newcommand{\eahnw}{$e\gamma \ra H \nu_e W$}

\newcommand{\thebb}{\theta_{b-beam}}
\newcommand{\thebc}{\theta_{b(c)-beam}}
\newcommand{\pte}{p^e_T}
\newcommand{\ptH}{p^H_T}
\newcommand{\gag}{$\gamma \gamma$ }
\newcommand{\gam}{\gamma \gamma }

\newcommand{\aatoh}{$\gamma \gamma \ra H \;$}
\newcommand{\egam}{$e \gamma \;$}
\newcommand{\eat}{e \gamma \;}
\newcommand{\eaeh}{$e \gamma \ra e H\;$}
\newcommand{\eaehb}{$e \gamma \ra e H \ra e (b \bar b)\;$}
\newcommand{\egebb}{$e \gamma (g) \ra e b \bar b\;$}
\newcommand{\egecc}{$e \gamma (g) \ra e c \bar c\;$}
\newcommand{\egebc}{$e \gamma (g) \ra e b \bar b(e c \bar c)\;$}
\newcommand{\eaebb}{$e \gamma \ra e b \bar b\;$}
\newcommand{\eaecc}{$e \gamma \ra e c \bar c\;$}
\newcommand{\aah}{$\gamma \gamma H\;$}
\newcommand{\zah}{$Z \gamma H\;$}
\newcommand{\pe}{P_e}
\newcommand{\pg}{P_{\gamma}}
\newcommand{\delbb}{\Delta m_{b \bar b}}
\newcommand{\delbc}{\Delta m_{b \bar b(c\bar c)}}


\title{\Large \bf {\boldmath \zah} vertex effects at future
        colliders
      }

\author{E.~Gabrielli$^a$, V.A.~Ilyin$^b$ and B.~Mele$^c$ \\ \\
         $^a$ University of Notre Dame, IN, USA \\
         $^b$ Institute of Nuclear Physics, Moscow State University, Russia \\
         $^c$ INFN, Sezione di Roma 1 and Rome University "La Sapienza", Italy}
\date{}

\maketitle

\vskip 0.5cm

The Higgs boson sector is a crucial part of the Standard Model (SM) still
escaping direct experimental verification. Once the scalar boson will be
discovered either at LEP2, upgraded TEVATRON or at LHC, testing its properties
will be a central issue at future linear  colliders. In particular, an \epem
collider with centre-of-mass (c.m.) energy $\sqrt{s}\simeq (300\div 2000)$GeV
and integrated luminosity ${\cal O}(100)$ fb$^{-1}$ will allow an accurate
determination of the mass, some couplings and  parity properties of this new
boson \cite{saar,zerw}. Among other couplings, the interaction of scalars  with
the neutral electroweak gauge bosons,  $\gamma$ and $Z$, are particularly
interesting. Indeed, one can hope to test here some delicate feature of the
Standard Model --- the relation between the spontaneous symmetry breaking
mechanism and the electroweak mixing of the two gauge groups $SU(2)$ and
$U(1)$. In this respect, three vertices could be measured --  $ZZH$,
$\gamma\gamma H$ and $Z\gamma H$. While the $ZZH$ vertex stands in SM at the
tree level, the other two contribute only at one-loop. This means that  the
\aah and \zah couplings could be sensitive to the contributions of new
particles circulating in the loop.

Here, we discuss the case of  an intermediate-mass Higgs boson, that is with
$M_Z\lta \mh\lta 140$ GeV.  A measurement of the \aah coupling should be
possible by the determination of the BR for the decay $H\ra \gamma \gamma$,
e.g. in the LHC Higgs discovery channel, $gg\to H\to\gamma\gamma$. 
Furthermore, at future photon-photon colliders\footnote{Two  further options
are presently considered for a high-energy $e^+e^-$ linear collider,  where one
or both the initial $e^+/e^-$ beams are replaced by   photon beams induced by
Compton backscattering of laser light on the high-energy electron beams
\cite{spec}. Then, the initial real photons could be to a good degree
monochromatic, and have energy and luminosity comparable to the ones of the
parent electron beam \cite{mono}.}, the precise measurement of the  \aah vertex
looks  realistic at the resonant production of the Higgs particle,
$\gamma\gamma\to H$. To this end, the capability of tuning the \gag c.m. energy
on the Higgs mass, through  a good degree of the photons  monochromaticity,
will be crucial for not diluting too much  the $\gam \to H$  resonant cross
section over the c.m. energy spectrum. Measuring the  \zah vertex is in general
more complicated.  Indeed, if one discusses the corresponding Higgs decay, the
final states include the $Z$ decay products, jets or lepton pairs, where much
heavier backgrounds are expected. Then, one can discuss  the $H\to \gamma Z$ 
decay only for $m_H\gta 115$ GeV (and $m_H\lta 140$),  when the corresponding
branching is as large as ${\cal O}(10^{-3})$.  Another possibility of measuring
the \zah vertex  is given by  collision processes. At electron-positron
colliders, the corresponding channels are $ e^+e^- \to \gamma H$ and $e^+e^-
\to Z H$.  However, in the $ZH$ channel the \zah vertex contributes to the
corresponding  one-loop corrections, thus implying a large tree level
background. The reaction $e^+e^-\to \gamma H$ has been extensively studied in
the literature \cite{barr,abba,djou}. Unfortunately, the \eeah channel suffers
from small rates, which are further depleted at large energies by the $1/s$
behavior of the dominant s-channel diagrams. For example, $\sigma_S\approx
0.05\div 0.001$ fb at $\sqrt{s}\sim 500\div 1500$ GeV.  We estimated the  main
background coming from the $e^+e^-\to\gamma b\bar b$ process, and found it
rather heavy: $\sigma_B\approx 4\div 0.8$ fb for $m_{b\bar b}=100\div 140$ GeV,
assuming a high resolution in the measurement of the  invariant  mass of $b\bar
b$ quark pair, i.e. $\pm 3$ GeV, and applying a minimum  cut of $18^\circ$  on
the angles [$\gamma-beams$] and [$b(\bar b)-beams$].   Then, at $\sqrt{s}=1.5$
TeV, one gets $\sigma_B\approx 0.4\div 0.07$ fb. One can conclude that
measuring the \zah vertex is not an easy task.  Recently, the Higgs production
in electron-photon collisions through the one-loop process   $e\gamma \to eH$
was analysed in details  \cite{noii}. This channel will turn out to be an
excellent mean to test  both the \aah and \zah one-loop couplings with high
statistics, without requiring a fine tuning of the c.m. energy. 

In this talk we discuss the prospects of the \eaeh reaction in setting
experimental bounds on the value of the anomalous \zah coupling. For this
analysis we use a model independent approach, where $dim=6 \;\;$ \su invariant
operators are added to the SM Lagrangian. In realistic  models extending the 
SM, these operators contribute in some definite combinations. However, if one
discusses the bounds on the possible deviations from  the standard-model
one-loop Higgs vertices, this approach can give some general  insight into the
problem. These anomalous operators contribute to all the three vertices   \aah,
\zah and $ZZH$, with only the first two involved in the \eaeh reaction.  As we
mentioned above, the anomalous contributions to the \aah vertex can be bounded
through the  resonant $\gamma\gamma\to H$ reaction, or by measuring the total
rate of the discussed reaction, \eaeh. Here we discuss the case where the
anomalous contributions to the \aah vertex have been well tested  in some other
experiment, and one would like to get limitations  just on the anomalous
contributions to the \zah vertex.   We underline that this specific case  is
out of the $\gamma\gamma$ domain. 

All the results presented in this talk were obtained with the help of CompHEP
package \cite{comp}. 


\vskip 0.3cm
{\bf \boldmath The reaction \eaeh: main features}

In \cite{noii}, we presented the complete analytical results for the helicity 
amplitudes of the \eaeh process (see also \cite{cott}). The total rate of this
reaction is rather high. For $\mh$ up to about 400 GeV, one finds $\sigma>1$
fb. 

The main strategy to enhance  the \zah vertex effects (depleted by the $Z$
propagator) with respect to the dominant \aah contribution\footnote{In
\cite{noii}, we define different contributions: `\aah' and `\zah', related to
the \aah and \zah vertices respectively, and a 'BOX' contribution.  The
separation of the rate into these three parts corresponds to the case  where
the Slavnov-Taylor identities for the `\aah' and `\zah' Green  functions just
imply the transversality with respect to the incoming photon momentum.}
consists in requiring a final electron  tagged at large angle. E.g., for
$\pte>100$GeV, \zah is about 60\% of \aah, and \zah gives a considerable
fraction of the total production rate, which is still sufficient to guarantee
investigation (about 0.7 fb).

The main irreducible background to the process $e\gamma\to eH\to eb\bar b$ 
comes from the channel \eaebb.  A further source of background is the charm
production through \eaecc, when the $c$ quarks are misidentified into $b$'s.
The cut $\thebc>18^\circ$ (between each $b(c)$ quark and both the
beams\footnote{we also assume a 10\% probability  of misidentifying a  $c$
quark into a $b$.})  reduces the signal and background at a comparable level.  
Numerically the \eaecc ``effective rate" is less than 1/3 of the \eaebb rate. A
further background, considered in \cite{noii}, is the resolved \egebc
production, where the photon interacts via its gluonic content. It was found
$<0.01$ fb at $\sqrt{s}=500$ GeV with our standard  cuts $p_T^e>100$ GeV and
$\thebc>18^\circ$.

We also considered the possibility of having the electron beam longitudinally
polarized. In table 1 results for the $e$ polarization dependence of  the total
cross section, its \aah, \zah and BOX components, and  their  interference 
pattern are collected for $m_H=120$ GeV.  This dependence turns out to be very
sensitive to the electron polarization. For instance, assuming $\pe=-1$
($\pe=+1$)  the total cross section increases (decreases) by about 94\% at
$\sqrt{s}=500$GeV. For $\pe=+1$, there is a strong destructive interference
between the terms \aah and \zah.

Important improvements in the $S/B$ ratio can be obtained by exploiting the
final-electron angular asymmetry in the signal. Indeed,  the final electron in
\eaeh moves mostly in the forward direction.   On the other hand, we found that
in the \eaebb background the final electron angular distribution, although not
completely symmetric, is almost equally shared in the forward and backward
direction with respect to the beam. In table~\ref{tab71}, we report, apart from
the total rate, the forward cross sections. It could also be convenient to
measure the FB asymmetry,  that has the advantage of being free from possible
uncertainties on the absolute normalization of the cross sections.

The main conclusion obtained in \cite{noii} is the following.  With a
luminosity of 100 fb$^{-1}$, at $\sqrt{s}=500$GeV, one expects an accuracy as
good as about 10\% on the measurement of the \zah effects assuming the validity
of Standard Model.  


\vskip 0.3cm
{\bf Anomalous vertices}
  
There are two pairs of $dim=6$ operators, CP-even and CP-odd respectively, 
giving anomalous contributions to the process \eaeh \cite{anomOP}:
$$ {\cal L}^{eff} = d\cdot {\cal O}_{UW} + d_B\cdot {\cal O}_{UB} +
     \bar d\cdot \bar {\cal O}_{UW} 
                        + \bar d_B\cdot \bar {\cal O}_{UB},$$
$$ {\cal O}_{UW} \,=\, \frac{1}{v^2} \left( |\Phi|^2-\frac{v^2}{2}\right)
    \cdot W^{i\mu\nu}  W^i_{\mu\nu}, \qquad
   {\cal O}_{UB} \,=\, \frac{1}{v^2} \left( |\Phi|^2-\frac{v^2}{2}\right)
    \cdot B^{\mu\nu}  B_{\mu\nu}, $$
$$ \bar {\cal O}_{UW} \,=\, \frac{1}{v^2}  |\Phi|^2
    \cdot W^{i\mu\nu}  \tilde W^i_{\mu\nu}, \qquad
   \bar {\cal O}_{UB} \,=\, \frac{1}{v^2} |\Phi|^2
    \cdot B^{\mu\nu}  \tilde B_{\mu\nu}, $$
where
$ \tilde W^i_{\mu\nu} = \epsilon_{\mu\nu\mu'\nu'}\cdot W^{i\mu'\nu'}$ and
$ \tilde B_{\mu\nu} = \epsilon_{\mu\nu\mu'\nu'}\cdot B^{\mu'\nu'}$.
In these formulas $\Phi$ is the Higgs doublet and 
$v$ is the electroweak vacuum expectation value.

The $Z\gamma H$ anomalous terms in the helicity amplitudes of $e\gamma\to eH$
are
$$ \frac{4\pi\alpha}{M_Z(M_Z^2-t)}  \sqrt{-\frac{t}{2}}
   \left\{ d_{\gamma Z} [(u-s)-\sigma\lambda (u+s)] 
   - i{\bar d_{\gamma Z}} [\lambda (u-s)+\sigma (u+s)] \right\},
$$
where $s$, $t$ and $u$ are the 
Mandelstam kinematical variables, $\sigma/2=\pm 1/2$
and $\lambda=\pm 1$ are the electron and photon  helicities, respectively. 
The
anomalous couplings contribute in the combinations $d_{\gamma Z}=d-d_B$ and
$\bar d_{\gamma Z}=\bar d-\bar d_B$.

Note that there is no interference between CP-odd terms (couplings $\bar d$)
and any triangle terms in the SM amplitude, although the interference  with the
BOX amplitude is nonvanishing.  This is due to the real value of the SM
triangle  amplitudes  for $M_H<2M_W$ and $M_H<2m_{top}$ (one can neglect
contributions of light quark loops), while the box diagrams contribute with
complex numbers. However,  the BOX amplitude gives a rather small contribution.
So, the cross section depends  only marginally on the sign of the CP-odd
couplings $\bar d_{\gamma Z}$. On the contrary, for CP-even terms, the
interference is relevant,  and the dependence on the $d_{\gamma Z}$ coupling is
not symmetrical with  respect to the SM point $d_{\gamma Z} =0$. This effect,
generally speaking, decreases the sensitivity to the CP violating anomalous
couplings.

\vskip 0.3cm
{\bf \boldmath Bounds on the anomalous \zah coupling}

As  we mentioned, we assume  that the anomalous contributions to the \aah
vertex are already tightly bounded somehow,  by using the data  from other
experiments, and one wants to get limitations on the anomalous contributions to
the \zah  vertex.  So, we discuss the case where the anomalous operators
contribute   in the combinations $d-d_B=d_{\gamma Z}$ and  $\bar d-\bar
d_B=\bar d_{\gamma Z}$ only. The bounds on the anomalous $d_{\gamma Z}$ and 
$\bar d_{\gamma Z}$ couplings presented below, have been computed by using  the
requirement that no deviation from the SM cross section is observed at the 
95\% CL:
$$ N^{anom}(\kappa) < 1.96 \cdot \sqrt{N^{tot}(\kappa)}, \quad 
   \kappa = d_{\gamma Z}, \bar d_{\gamma Z}\;,   $$
$$ N_{tot}(\kappa) = {\cal L}_{int} \cdot [\sigma_S(\kappa)+\sigma_B]\;,
  \quad
   N^{anom}(\kappa) = {\cal L}_{int} \cdot 
     [\sigma_S(\kappa)-\sigma_S(0)]\;.
  $$
Here, by $\sigma_S(\kappa)$ we denote the cross  section of the signal reaction
$e\gamma\to eH\to eb\bar b$ with the anomalous contributions, so  $\sigma_S(0)$
is the standard-model cross section.  Then, by $\sigma_B$ we denote the total
cross section of the  background processes  \eaebb, \eaecc (with 10\%
probability of misidentifying a $c$ quark into a $b$ quark), and the
corresponding contributions from the resolved photons. ${\cal L}_{int}$ is the
integrated luminosity. The results of these calculations are collected in the
last two columns in  table~\ref{tab71}, where we present the bounds on the
anomalous couplings $d_{\gamma Z}$ and $\bar d_{\gamma Z}$, attainable with  an
integrated luminosity of 100 fb$^{-1}$. We find, that these bounds do not
depend significantly on the electron angular cut while the $S/B$ ratio is
improved up to $\sim 1$. Nevertheless, we stress that the measurement of  the
relative final-electron angular asymmetry could be more convenient, if the
systematic errors connected with the absolute normalization of the cross
sections are relevant. Furthermore, one can see from the table ~\ref{tab71},
that the  strongest limitation on the CP-even coupling (at the level
$|d_{\gamma Z}|\lta 0.0015$) is  obtained from a left-handed polarized electron
beam.  On the other hand,  in the CP-odd case, the right-handed polarized
electron beam gives the best performance, with bounds at the level of $|\bar
d_{\gamma Z}|\lta 0.005$. In the latter case, the violation of the strong
destructive interference between the \zah and \aah terms by the anomalous terms
compensates the decrease in statistics. As a result, we get better bounds. Even
in the CP-even case, a right-handed polarized electron beam could add some
valuable information to the `left-handed' data. Indeed, due to the rather large
interference between the CP-even anomalous terms with the SM amplitudes, the
experimental data (obtained at integrated luminosity 100 fb$^{-1}$) will give
two limitation intervals. One interval includes the SM point $d_{\gamma Z}=0$,
and the other a disconnected range $0.019<d_{\gamma Z}<0.022$. This is due to
the asymmetry of  the cross section with respect to the SM point $d_{\gamma
Z}=0$. Hence, an experiment with   a right-handed polarized $e$ beam
could  resolve these {\it discrete} uncertainty.

Finally, let us rewrite the obtained bounds on the anomalous couplings 
in terms of
the mass scale of the {\it New Physics}. 
We introduced the anomalous couplings as
dimensionless parameters in the effective Lagrangian using the electroweak
scale $v\approx 250$ GeV. If these anomalous terms appear as contributions 
of new particles in the \zah loop with the mass $M_{new}$, then one gets
$d_{\gamma Z},\bar d_{\gamma Z}\sim (v/M_{new})^2$. By using this
relation, one obtains the bounds $ M_{new}>6.2$ TeV in the
CP-even case and $ M_{new}>3.5$ TeV in the CP-odd case.

The participation of V.A.I. to this Conference has been supported by the
Org. Committee  and by the Russia Ministry for Science and Technologies.



\begin{table}[bh]
\caption{Interference pattern between the $\gamma\gamma Z$,
$Z\gamma H$ and boxes contributions versus the 
$e$-beam polarizations, for 
$p^e_T > 100\,$GeV. $\gamma$-beam is unpolarized.}
\label{tab53b}
\begin{tabular}{|c||c|c|c|c|c|c|c|}
\hline
       &\multicolumn{7}{c|}{$\sigma(e\gamma\to eH)$ fb} \\
                 \cline{2-8}
  $m_H=120$ GeV 
 & & & & 
     & \multicolumn{3}{c|}{Interference terms} \\
                 \cline{6-8}
 $\sqrt{s}=500$ GeV  
     & Total & $|\gamma\gamma H|^2$& $|Z\gamma H|^2$ & $|$BOX$|^2$
    & \small ${\gamma\gamma H -}$  & \small ${\gamma\gamma H -}$  
                      & \small ${Z\gamma H -}$ \\
 & & & & & \small ${Z\gamma H}$  & \small ${BOX}$ 
                      & \small ${BOX}$ \\
\hline
 $P_e = 0$ &    
     0.705  &   0.296  &   0.166  &   0.0532 &  0.0439 & 0.0766 & 0.0692  \\  
\hline
 $P_e = -1$ &
      1.37  &   0.296  &   0.199  &   0.106  &  0.478  & 0.147  & 0.144 \\
\hline
 $P_e = +1$ &
     0.0405 & 0.296  &   0.133   &  0.0004 & -0.390  &   0.0067& -0.0057 \\
\hline
\end{tabular}
\end{table}

\begin{table}[bh]
\caption{Bounds on the anomalous \zah couplings for different
$e$-beam polarizations.
The cross sections are shown for signal process
and for background, where
all contributions are collected, from $e\gamma\to
eb\bar b$ and $e\gamma\to ec\bar c$ (assuming 10\% of misidentifying a $c$ quark
into a $b$) and from the resolved photons.
The cuts ${\protect\theta_{b-beam}>18^\circ}$, $p_t^e>100$ GeV and 
${\protect\mh-3\mbox{GeV}< \mbc <\mh+3\mbox{GeV}}$ 
are applied.}
\label{tab71}
\begin{tabular}{|c||c|c||c|c|}
\hline
$m_H=120$ GeV& \multicolumn{2}{|c||}{$ \sigma_S/\sigma_B$ fb}
             & \multicolumn{2}{|c|}{Bounds at 100 fb$^{-1}$} \\
\cline{2-5}
\raisebox{0ex}[3ex][2ex]{$\sqrt{s}=500$ GeV}
&no $\theta_e$ cut&$\theta_e<90^\circ$
&$d_{\gamma Z}$&$\bar d_{\gamma Z}$ \\
\hline
$P_e = 0$ & 0.530/0.881 & 0.425/0.483&$(-0.0025,0.004)$&$(-0.006,0.006)$\\
$P_e = -1$& 1.03/1.307  & 0.820/0.719&$(-0.0015,0.0015)$&$(-0.006,0.006)$\\
$P_e = 1$ & 0.0249/0.472& 0.0245/0.270&$(-0.007,0.004)$&$(-0.005,0.005)$\\
\hline
\end{tabular}
\end{table}

\end{document}